\documentclass[amsmath,amssymb,aps,prx,reprint,superscriptaddress,longbibliography,twocolumn]{revtex4-2}

\usepackage{graphicx}% Include figure files
\usepackage[colorlinks, linkcolor=blue, urlcolor=blue, anchorcolor=blue, citecolor=blue]{hyperref}
\usepackage{color}
\usepackage{stmaryrd}
\usepackage{upgreek}

\makeatletter
\newcommand{\doublewidetilde}[1]{{%
  \mathpalette\double@widetilde{#1}%
}}
\newcommand{\double@widetilde}[2]{%
  \sbox\z@{$\m@th#1\widetilde{#2}$}%
  \ht\z@=.9\ht\z@
  \widetilde{\box\z@}%
}
\makeatother

\begin{document}

\title{Performance Analysis for Crosstalk Errors between Parallel Entangling Gates in Trapped Ion Quantum Error Correction}

\author{Fangxuan Liu}
\thanks{These authors contribute equally to this work}%
\affiliation{Center for Quantum Information, Institute for Interdisciplinary Information Sciences, Tsinghua University, Beijing 100084, P. R. China}
\affiliation{Shanghai Qi Zhi Institute, AI Tower, Xuhui District, Shanghai 200232, China}

\author{Gaoxiang Tang}
\thanks{These authors contribute equally to this work}%
\affiliation{Center for Quantum Information, Institute for Interdisciplinary Information Sciences, Tsinghua University, Beijing 100084, P. R. China}
\affiliation{Shanghai Qi Zhi Institute, AI Tower, Xuhui District, Shanghai 200232, China}

\author{Luming Duan}
\affiliation{Center for Quantum Information, Institute for Interdisciplinary Information Sciences, Tsinghua University, Beijing 100084, P. R. China}
\affiliation{Hefei National Laboratory, Hefei 230088, P. R. China}
\affiliation{New Cornerstone Science Laboratory, Beijing 100084, PR China}

\author{Yukai Wu}
\email{wyukai@mail.tsinghua.edu.cn}
\affiliation{Center for Quantum Information, Institute for Interdisciplinary Information Sciences, Tsinghua University, Beijing 100084, P. R. China}
\affiliation{Hefei National Laboratory, Hefei 230088, P. R. China}
\affiliation{Shanghai Qi Zhi Institute, AI Tower, Xuhui District, Shanghai 200232, China}

\begin{abstract}
The ability to execute a large number of quantum gates in parallel is a fundamental requirement for quantum error correction, allowing an error threshold to exist under the finite coherence time of physical qubits. Recently, two-dimensional ion crystals have been demonstrated as a plausible approach to scale up the qubit number in a trapped ion quantum computer. However, although the long-range Coulomb interaction between the ions enables their strong connectivity, it also complicates the design of parallel gates and leads to intrinsic crosstalk errors. Here we examine the effects of crosstalk errors on a rotated surface code. We show that, instead of the distance-3 code considered in previous works, a distance-5 code is necessary to correct the two-qubit crosstalk error. We numerically calculate the logical error rates and coherence times under various crosstalk errors, gate infidelities and coherence times of the physical qubits, and we optimize the parallelism level according to the competition between different error sources. We show that a break-even point can be reached under realistic parameters. We further analyze the spatial dependence of the crosstalk, and discuss the scaling of the logical error rate versus the code distance for the long-term goal of a logical error rate below $10^{-10}$.
\end{abstract}

\maketitle

\section{Introduction}
Ion trap is one of the leading physical platforms for quantum computation \cite{10.1063/1.5088164}, with the highest fidelity for state initialization and readout \cite{Crain2019,Roman_2020,PhysRevA.104.012606,PhysRevLett.129.130501}, and single-qubit \cite{PhysRevLett.113.220501,smith2024single} and two-qubit \cite{PhysRevLett.117.060504,PhysRevLett.117.060505,PhysRevLett.127.130505,loschnauer2024scalable} quantum gates. Tens of individually addressed ionic qubits have been realized in one-dimensional (1D) ion crystals \cite{Chen2024benchmarkingtrapped,Postler2022} and in quantum charge-coupled device (QCCD) architecture \cite{PhysRevX.13.041052,decross2024computational,wineland1998experimental,Kielpinski2002}. However, due to the experimental noise, the number of ions that can be stably trapped in the commonly used 1D structure is usually restricted to below 100 \cite{wineland1998experimental,PhysRevLett.77.3240,clark2001proceedings}. Although the QCCD architecture can surpass this limitation by dividing a large number of ions into nearly isolated small groups, it is currently bottlenecked by the relatively slow ion transportation and sympathetic cooling processes \cite{PhysRevX.13.041052}. Recently, two-dimensional (2D) ion crystals \cite{Szymanski2012crystal,https://doi.org/10.1002/qute.202000068,Xie_2021,PhysRevA.105.023101,PRXQuantum.4.020317,Qiao2024,guo2024siteresolved} have become a promising alternative approach to scale up the qubit number, demonstrating the stable trapping of up to 512 ions and the global quantum control of up to 300 ions \cite{guo2024siteresolved}. Individually addressed quantum gates on 2D ion crystals have also been reported, entangling arbitrary ion pairs in a small 4-ion crystal \cite{Hou2024}.

For quantum computation on a large ion crystal, the long-range Coulomb interaction between the ions enables their strong connectivity \cite{10.1063/1.5088164}. Direct entangling gates between arbitrary ion pairs have been achieved for 30 ions in 1D \cite{Chen2024benchmarkingtrapped}, which can largely reduce the circuit depth when compiling quantum algorithms compared with nearest-neighbor connectivity. However, such a long-range interaction also leads to crosstalk errors on parallel entangling gates \cite{Figgatt2019,Lu2019}. Because the capability to execute gates in parallel lays one of the foundations for quantum error correction (QEC) to timely remove the idling errors on individual qubits before they accumulate \cite{nielsen2000quantum}, it is crucial to suppress such crosstalk errors and to incorporate them into the theory of fault-tolerant quantum computation on trapped ions. On the one hand, polynomial-time algorithms have been developed to design the pulse sequence for parallel entangling gates on arbitrary ion pairs in a large ion crystal \cite{Grzesiak2020}, with the crosstalk error between undesired ion pairs set to zero in the ideal case through the cancellation between different pulse segments. However, in practice any noise in the experimental parameters like the laser frequencies and amplitudes can prevent the perfect cancellation and lead to nonzero crosstalk errors. On the other hand, various schemes have been proposed to suppress the crosstalk in a more error-robust way through the engineering of the collective phonon modes of the ion crystal by applying external optical tweezers \cite{PRXQuantum.1.020316,PhysRevApplied.22.034021} or introducing different ion species \cite{PhysRevA.108.042603}. However, these schemes complicate the ion trap setup and can be challenging for large ion crystals.

In this work, we analyze the effects of the crosstalk errors on ion trap quantum error correction. Note that throughout this paper by crosstalk we mean the undesired spin-spin entanglement due to the coupling to the shared collective phonon modes \cite{Figgatt2019,PRXQuantum.1.020316,PhysRevApplied.22.034021,PhysRevA.108.042603}. This should be distinguished from another type of crosstalk considered in previous literatures due to a finite laser beam radius compared with the ion distance \cite{PhysRevA.88.052325,Debroy_2020,ParradoRodriguez2021crosstalk,PhysRevLett.129.240504,10.1063/5.0177638,flannery2024physical}. In those works the crosstalk errors are local and coherent, allowing potential cancellation via the spatial refocusing technique \cite{PhysRevA.88.052325} or simply spin echoes \cite{Grzesiak2020,PhysRevLett.129.240504}. In comparison, as we will show, in our case the crosstalk errors are long-range two-qubit errors, necessitating the use of distance-5 QEC codes rather than the commonly used distance-3 codes for a pseudothreshold to exist.

The paper is organized as follows. In Sec.~\ref{sec:background} we introduce the necessary background, including the rotated surface code in Sec.~\ref{sec:code} and the physical model of the parallel entangling gates in ion trap and their crosstalk in Sec.~\ref{sec:error_model}.
In Sec.~\ref{sec:distance5} we present the detailed analysis for the performance of a $\llbracket 25, 1, 5\rrbracket$ rotated surface code \cite{bravyi1998quantumcodeslatticeboundary,PhysRevA.76.012305,PhysRevA.86.032324} with $24$ ancilla qubits for stabilizer measurements. We calculate the logical error rates and the logical qubit coherence times under practical gate infidelities, crosstalk errors and physical qubit decoherence rates, and we discuss parameter regimes where a break-even point can be achieved. Then in Sec.~\ref{sec:optimal_parallelism} we further optimize the parallelism level of the entangling gates and show the competition between the crosstalk error and the idling error. In Sec.~\ref{sec:scaling} we discuss the scalability of ion trap QEC under the crosstalk error based on detailed examples of parallel entangling gates on 2D ion crystals. We consider two parameter regimes: For a gate speed slower than the spatial propagation of local phonon excitations, the crosstalk error is long-range and nearly independent of the distance between the parallel gates; For a gate speed faster than the acoustic speed on the 2D lattice, the crosstalk is local and decays polynomially with the distance between gate pairs. In both cases, we discuss how a logical error rate below $10^{-10}$ can be reached, which is a long-term goal for many practical applications of quantum computation \cite{PhysRevA.86.032324,Acharya2024}, using a 2D crystal with hundreds to thousands of ions. Appendix~\ref{sec:appendix} provides numerical and analytical results for a unified expression of the logical error rate under the gate infidelity, idling error and crosstalk error. Appendix~\ref{sec:EASE} reviews an efficient algorithm, the EASE gate protocol \cite{Grzesiak2020}, to design parallel entangling gates in a large ion crystal.

\begin{figure}[!tbp]
    \centering
    \includegraphics[width=\linewidth]{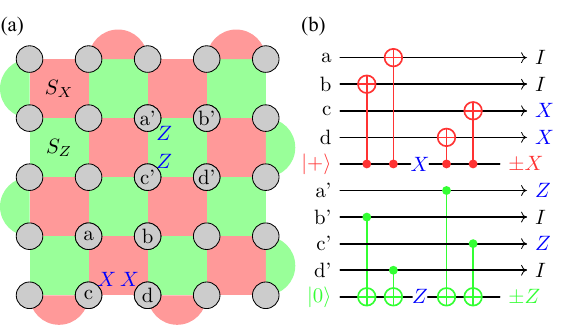}\\
    \caption{(a) An illustration of a $d=5$ rotated surface code. The grey circles denote data qubits, and the red (green) plaquettes represent the $X$ ($Z$) stabilizers, each with an ancilla qubit for measurement (not shown in the plot). The blue symbols indicate one possible pattern of errors on the data qubits after one round of syndrome measurement due to the occurrence of a two-qubit crosstalk error. (b) Quantum circuits for the measurement of $X$ (top) and $Z$ (bottom) stabilizers. CNOT gates in the same layer can be applied in parallel. Suppose a crosstalk error occurs between two CNOT gates in the second layer, leading to a two-qubit $XZ$ error (blue) on two ancilla qubits. After the following gates, this crosstalk error propagates into a weight-4 error on the data qubits. Fortunately, the scheduling of the CNOT gates in the surface code ensures that this weight-4 error can still be corrected with a code distance $d\ge 5$. \label{fig:surface}}
\end{figure}

\section{Background}
\label{sec:background}
\subsection{Rotated surface code}
\label{sec:code}
The rotated surface code is a family of $\llbracket d^2,1,d \rrbracket$ QEC code where one logical qubit is encoded into $d^2$ data qubits to get protection from all the possible errors acting on less than or equal to $\lfloor (d-1)/2 \rfloor$ physical qubits. An example of a $d=5$ rotated surface code is shown in Fig.~\ref{fig:surface}(a). Apart from the $d^2$ data qubits, it further uses $d^2-1$ ancilla qubits to extract the error syndrome from the $X$-type and $Z$-type stabilizers. These stabilizers are indicated by the red (green) plaquettes in Fig.~\ref{fig:surface}(a) as the tensor product of Pauli $X$ ($Z$) operators on the data qubits surrounding the plaquettes.

As a variant of the surface code using less physical qubits \cite{bravyi1998quantumcodeslatticeboundary,PhysRevA.76.012305,PhysRevA.86.032324}, the rotated surface code shares the advantage of a high threshold, which is preferable for near-term demonstrations, and requires only nearest-neighbor operations on a 2D lattice for error correction, leaving it a promising candidate for quantum computation platforms like superconducting circuits. As we will show later, although ion trap naturally supports long-range entangling gates, this local property can still be utilized to suppress the effect of crosstalk errors.

Furthermore, the measurement of all the stabilizers in the rotated surface code can be performed in parallel, making it an excellent target for our purpose to study the crosstalk between parallel gates. Each stabilizer measurement involves at most four CNOT gates between an ancilla qubit and four neighboring data qubits. These CNOT gates can be parallelized into four layers as shown in Fig.~\ref{fig:surface}(b), each containing $d(d-1)$ pairs of gates. This results in a constant circuit depth independent of the code distance $d$. For stabilizers on the boundary of the code, only two neighboring data qubits are involved, and we can simply remove two corresponding CNOT gates in the circuit.
Note that the parallel execution of the stabilizer measurements is crucial for QEC under a finite coherence time of the physical qubits. Later in Sec.~\ref{sec:optimal_parallelism} we will examine the competition between the increasing crosstalk error and the decreasing idling error when parallelizing more gates and discuss the optimal parallelism level.

To make the QEC process fault-tolerant, the order of the CNOT gates in Fig.~\ref{fig:surface}(b) are carefully designed \cite{PhysRevA.86.032324}. The goal is to prevent the propagation of errors in the direction of the logical operators. For example, as shown in Fig.~\ref{fig:surface}(b), a single-qubit $X$ ($Z$) error propagates in the horizontal (vertical) direction, perpendicular to the logical $X$ ($Z$) operator which is a weight-5 vertical (horizontal) Pauli $X$ ($Z$) string across the 2D lattice. In this work, apart from the commonly considered single-qubit and two-qubit gate errors, we further include two-qubit crosstalk errors between the parallel entangling gates. As an example, in Fig.~\ref{fig:surface}(b) we may have a correlated $XZ$ error on two ancilla qubits during the second layer of the entangling gates, which further propagates into a weight-4 error on the data qubits. Fortunately, owing to the scheduling of the CNOT gates described above, such errors can still be corrected by a code with a distance $d\ge 5$.

Given the outcome of at least $d$ rounds of syndrome measurement, we can use the minimum-weight perfect matching (MWPM) algorithm \cite{Edmonds_1965,edmonds1965maximum} to decode the error syndrome. In this work we use the Stim package \cite{Gidney2021stimfaststabilizer} to numerically simulate the occurrence of the physical errors and the performance of the QEC code for a logical qubit initialized in the logical $|0_L\rangle$ state.

\subsection{Parallel gates in ion trap and the noise model}
\label{sec:error_model}
In ion trap, high-fidelity entangling gates are usually realized by the Molmer-Sorensen gate scheme \cite{Sorensen1999,PhysRevLett.82.1835} or the light shift gate scheme \cite{milburn2000,leibfried2003experimental} based on the spin-dependent force. Their mathematical frameworks are similar and the difference mainly lies in the how the spin-dependent force is generated and whether the gate is in the $X$ or $Z$ basis. Here we consider the Molmer-Sorensen gate as an example, which has been realized for ${}^{171}\mathrm{Yb}^+$ ions on a 2D lattice by pulsed $355\,$nm laser beams \cite{Hou2024}.

We start from the spin-dependent-force Hamiltonian \cite{Zhu2006}
\begin{equation}
H= \sum_{jk} \eta_k b_{jk}\Omega_j \sin \mu t \left(a_k e^{-i\omega_k t} + a_k^\dag e^{i\omega_k t}\right) X_j,
\end{equation}
where the index $k$ labels the collective phonon modes of the ion crystal with the frequency $\omega_k$, the Lamb-Dicke parameter $\eta_k$, and the annihilation (creation) operator $a_k$ ($a_k^\dag$), and the index $j$ enumerates the target ions, each addressed by a controllable laser amplitude $\Omega_j$. The coefficients $b_{jk}$ represent the participation of each ion $j$ in the phonon mode $k$, and $X_j$'s are the Pauli $X$ operators. Here we set the laser phase to zero and use $\mu$ to denote the laser detuning from the resonant transition of the qubits.

Note that the above Hamiltonian preserves the $X_j$ operators. Therefore, for each spin state in the $X$ basis, the time evolution is just displacements of all the phonon states, from which a geometric phase can appear due to the commutation relation between displacements in different directions. The overall unitary evolution after a gate time $\tau$ can be described by two-qubit entanglements between all the ion pairs and spin-dependent displacements on each phonon mode \cite{Zhu2006}
\begin{equation}
U(\tau) = \exp\left(i \sum_{i<j}\Theta_{ij}X_i X_j\right)\prod_{k} D_k\left(\sum_j\alpha_{jk} X_j\right),
\end{equation}
where
\begin{equation}
D_k(\alpha) = \exp(\alpha a_k^\dag - \alpha^* a_k)
\end{equation}
is the displacement operator of the $k$-th mode,
\begin{equation}
\alpha_{jk} = -i \eta_k b_{jk} \int_0^\tau \chi_j(t) e^{i\omega_k t} dt \label{eq:alpha}
\end{equation}
represents the residual spin-phonon entanglement after the gate,
\begin{align}
\Theta_{ij}=&\sum_k \eta_k^2 b_{ik} b_{jk} \int_0^\tau dt_1 \int_0^{t_1} dt_2 \nonumber\\
&\left[\chi_i(t_1) \chi_j(t_2) + \chi_j(t_1) \chi_i(t_2)\right]\sin \left[\omega_k (t_1-t_2)\right] \label{eq:theta}
\end{align}
characterizes the spin-spin entanglement between all the ion pairs, and
\begin{equation}
\chi_j(t) = \Omega_j \sin\mu t \label{eq:laser}
\end{equation}
describes the controllable laser driving on the target ions.

To realize a maximally entangled two-qubit gate, ideally we want $\alpha_{jk}=0$ for both ions and all the phonon modes, and $\Theta_{ij}=\pm\pi/4$ between the ion pair. This can be realized by a suitable sequence of the driving laser in Eq.~(\ref{eq:laser}) by controlling its amplitude, frequency or phase \cite{Zhu2006,PhysRevLett.114.120502,PhysRevA.98.032318}. This maximally entangled gate is equivalent to a CNOT gate up to additional single-qubit rotations \cite{10.1063/1.5088164}. Similarly, here to parallelize multiple entangling gates, we want all the $\alpha_{jk}$'s to vanish, while all the $\Theta_{ij}$'s to take the desired values $\theta_{ij}$ for parallel entangling gates. An efficient numerical algorithm to design the laser pulse sequence by amplitude modulation is the EASE algorithm \cite{Grzesiak2020}, which is briefly summarized in Appendix~\ref{sec:EASE}.

Small deviation from the above ideal conditions can lead to an average gate infidelity \cite{PhysRevA.97.062325}
\begin{equation}
\delta F = \frac{D}{D+1} \left[\sum_{jk}|\alpha_{jk}|^2(2\bar{n}_k+1) + \sum_{i<j}(\Theta_{ij}-\theta_{ij})^2\right],
\end{equation}
where $\bar{n}_k$ is the average thermal phonon number of the mode $k$ and $D=2^N$ is the total dimension of the system with $N$ qubits.
Below we analyze how this total infidelity decompose into single-qubit and two-qubit errors.

Clearly, nonzero $\Theta_{ij}$'s for undesired ion pairs lead to $XX$-type crosstalk errors, while deviations of $\Theta_{ij}$'s from the ideal values of $\pm \pi/4$ for the targeted entangling gates contribute to part of the two-qubit gate errors. These deviations may arise from the imperfect pulse design or miscalibration of parameters in the experiment. Note that given an imperfect pulse sequence or given the pattern of parameter drifts in each experimental shot, these errors may be coherent, which is detrimental for quantum error correction. For example, we may obtain an additional $e^{i\epsilon_{ij} X_i X_j}$ gate ($\epsilon_{ij}\ll 1$) between an undesired ion pair as a crosstalk error. Fortunately, such coherent errors can be turned into incoherent ones by inserting randomized single-qubit rotations \cite{Kern2005,PhysRevA.94.052325}. Therefore, here we model the above crosstalk error as a two-qubit Pauli $XX$ error, where an $X_i X_j$ gate is applied with a probability $p_c=\epsilon_{ij}^2$. Note that after compiling the CNOT gates used in QEC into the Molmer-Sorensen gates, this $XX$ error will be transformed accordingly by the single-qubit gates. As shown in Fig.~\ref{fig:surface}(b), we always have $X$-type errors on the control qubits and $Z$-type errors on the target qubits.

As for the residual spin-dependent displacement term, we trace out the phonon states to examine its effect on the qubits. Following the commonly used assumption, we consider an initial thermal state of the phonons $\rho_m$ with an average phonon number $\bar{n}_k$ in each mode $k$. Then an initial spin state $\rho$ evolves into
\begin{equation}
\mathrm{Tr}_m \!\!\left[\prod\nolimits_{k} \!\!D_k\left(\sum\nolimits_j\!\!\alpha_{jk}X_j\right) \rho\otimes \rho_m \prod\nolimits_{l}\!\! D_l^\dag\left(\sum\nolimits_i\!\!\alpha_{il}X_i\right)\right].
\end{equation}

We can consider a specific matrix element $(s,s^\prime)$ in the $X$ basis, where $s$ and $s^\prime$ represent two vectors of spin states $s_j,s_i^\prime \in \{+1,-1\}$. The evolved matrix element is
\begin{align}
& \rho_{s,s^\prime}\mathrm{Tr} \left[\prod\nolimits_{k} \!\!D_k\left(\sum\nolimits_j\!\!\alpha_{jk}s_j\right) \rho_m \prod\nolimits_{l}\!\! D_l^\dag\left(\sum\nolimits_i\!\!\alpha_{il}s_i^\prime\right)\right]\nonumber\\
\approx &\rho_{s,s^\prime}\mathrm{Tr} \left\{\prod\nolimits_{k} D_k\left[\sum\nolimits_j\alpha_{jk}(s_j-s_j^\prime)\right] \rho_m \right\} \nonumber\\
=& \rho_{s,s^\prime}\exp\left[-\sum\nolimits_{k} \left|\sum\nolimits_j\alpha_{jk}(s_j-s_j^\prime)\right|^2 (\bar{n}_k+1/2) \right], \label{eq:trace}
\end{align}
where in the second line we have dropped a phase factor when combining two displacement operators in different directions. The dropped phase is on the order of $O(|\alpha|^2)$, corresponding to an infidelity of $O(|\alpha|^4)$ which is dominated by the remaining term on the order of $O(|\alpha|^2)$.

A direct observation is that Eq.~(\ref{eq:trace}) represents a dephasing error in the $X$ basis. For the diagonal matrix elements $s=s^\prime$ we have the terms inside the bracket equal to zero, while the off-diagonal elements of the density matrix are reduced according to the residual spin-dependent displacement. In general, this dephasing is a joint effect on all the spins and needs to be described as a multi-qubit error. However, if we further assume that different $\alpha_{jk}$'s are random numbers with independent magnitudes and directions, then the effect of the residual spin-dependent displacement can be represented as
\begin{align}
\rho_{s,s^\prime} \to& \rho_{s,s^\prime}\exp\left[-\sum\nolimits_{jk} \left|\alpha_{jk}\right|^2(s_j-s_j^\prime)^2 (\bar{n}_k+1/2) \right] \nonumber\\
=& \rho_{s,s^\prime}\prod\nolimits_j\!\!\exp\left[-(s_j-s_j^\prime)^2\sum\nolimits_{k} \left|\alpha_{jk}\right|^2 (\bar{n}_k+1/2) \right],
\end{align}
which is just a dephasing error for each spin individually. This error model ensures that, no matter how many gate pairs we apply in parallel, we only need to deal with single-qubit and two-qubit errors, then we can expect the errors to be correctable by increasing the code distance.
To justify the required assumption, note that ideally we have $\alpha_{jk}=0$ and the nonzero values can arise from independent miscalibration of laser pulses on individually addressed ions. Also note that we may introduce random laser phases for different pairs of parallel gates, which on the one hand suppress their crosstalk $\Theta_{ij}$'s, and on the other hand also remove the correlation between the dephasing terms.
Given the above analysis for the two-qubit nature of the gate error, we use the standard two-qubit depolarizing channel to model the two-qubit gate error, with an error probability of $p_g$ for each targeted entangling gate.

Finally, due to the importance of the competition between the crosstalk error and the idling error for parallel gates, we also model an independent single-qubit idling error for all the physical qubits. We take the duration of a two-qubit entangling gate as the time unit, and set the typical duration of a single-qubit gate as $0.1$ and that of the state detection as $5$. In the following sections, we either set a physical coherence time $T$ such that any operation with a duration of $t$ is subjected to a single-qubit depolarizing error with probability $p_i=(3/4)(1-e^{-t/T})$, where the factor of $3/4$ is the maximal depolarizing error to generate a fully mixed state, or we report an idling error rate $p_i$ for an operation with a unit duration, from which the physical coherence time can be computed as
\begin{equation}
T=-t_0/\ln[1-(4/3)p_i], \label{eq:coherence}
\end{equation}
where we set the unit duration $t_0=1$.

\begin{figure*}[!tbp]
    \centering
    \includegraphics[width=\linewidth]{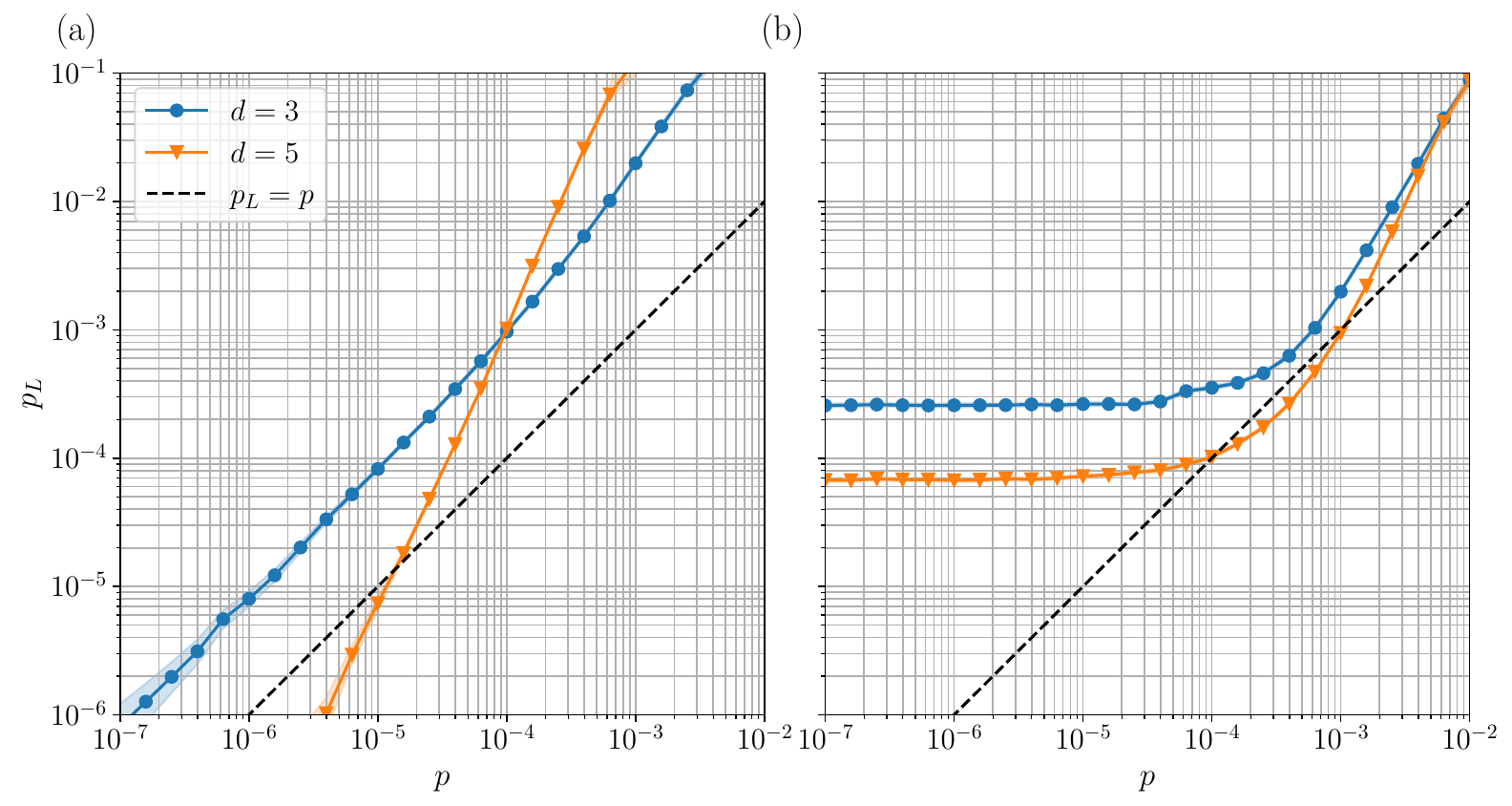}\\
    \caption{Logical error rates of $d=3$ (blue) and $d=5$ (orange) codes under physical errors including the two-qubit crosstalk. (a) We assume the same value $p$ for the two-qubit gate infidelity $p_g$, the idling error on physical qubits $p_i$, and the crosstalk between each pair of parallelized two-qubit gates $p_c$. For $d=3$, we observe a logical error rate per each round of syndrome measurement $p_L \propto p$ for sufficiently small physical error rate $p$, and we always have $p_L>p$. For $d=5$, we have $p_L \propto p^2$ and a pseudo-threshold exists where the logical error rate becomes smaller than the physical error rate. (b) We fix $p_c=10^{-4.5}$ and vary $p_g=p_i=p$. Under such a practical crosstalk error rate, a pseudo-threshold does not exist for the $d=3$ code no matter how we suppress the gate infidelity and the idling error, while it can still exist for the $d=5$ code. \label{fig:d_compare}}
\end{figure*}

\section{Correct crosstalk errors by distance-5 code}
\label{sec:distance5}
From the error model described in Sec.~\ref{sec:error_model}, we can see that the crosstalk errors are two-qubit errors and generally do not have a well-designed spatial pattern as the two-qubit gate errors. Therefore, these crosstalk errors cannot be corrected by distance-3 codes like the Steane code or the $d=3$ rotated surface code, and we need a minimal code distance $d=5$ to demonstrate the correction of such crosstalk errors.

This fundamental difference between $d=3$ and $d=5$ rotated surface codes can be seen in Fig.~\ref{fig:d_compare}(a). Here we set the same value $p$ for the two-qubit gate infidelity $p_g$, the idling error $p_i$ and the crosstalk error $p_c$, and simulate the logical error rate per one round of syndrome measurement $p_L$. For $d=3$, one crosstalk error is sufficient to cause an incorrect prediction of the decoder as its error syndrome can be indistinguishable from a single-qubit error. Indeed, we get $p_L \propto p$, and the logical error rate is always above the physical error rate by about one order of magnitude. This is because we are parallelizing $k=d(d-1)=6$ pairs of gates. Since there are $2k(k-1)$ possible locations for the two-qubit crosstalk error, effectively the crosstalk error per targeted entangling gate is scaled into $\tilde{p}_c=2(k-1)p_c=10p_c$. In contrast, for the $d=5$ code, a single crosstalk error, as well as a gate error or an idling error, can always be corrected. Hence we observe $p_L \propto p^2$ in the small-$p$ regime with a pseudo-threshold occurring around $p \approx 1.3\times 10^{-5}$ where the logical error rate matches the physical error rate. As we will see later, this level of crosstalk error is achievable for parallel gate design under practical noise.

In the above definition of the pseudo-threshold, we compare the logical error rate $p_L$ to all the physical error rates including $p_g$, $p_i$ and $p_c$. However, as the crosstalk error is usually not included in previous definitions \cite{PhysRevA.90.062320}, here we may also want to separate it from the rest of the physical errors. As shown in Fig.~\ref{fig:d_compare}(b), we set $p_i=p_g=p$, and choose a practical crosstalk error $p_c=10^{-4.5}\approx 3\times 10^{-5}$ to maximize the difference between the $d=3$ and the $d=5$ codes. Here no matter how we suppress the gate infidelity and the idling error, the logical error rate will always be lower bounded by some values governed by $p_c$, which is $O(p_c)$ for the $d=3$ code and $O(p_c^2)$ for the $d=5$ code. For the chosen value of $p_c$, we can observe that a pseudo-threshold does not exist for the $d=3$ code, but can still be reached for a reasonable gate infidelity and idling error around $p=10^{-3}$.

\begin{figure*}[!tbp]
    \centering
    \includegraphics[width=0.8\linewidth]{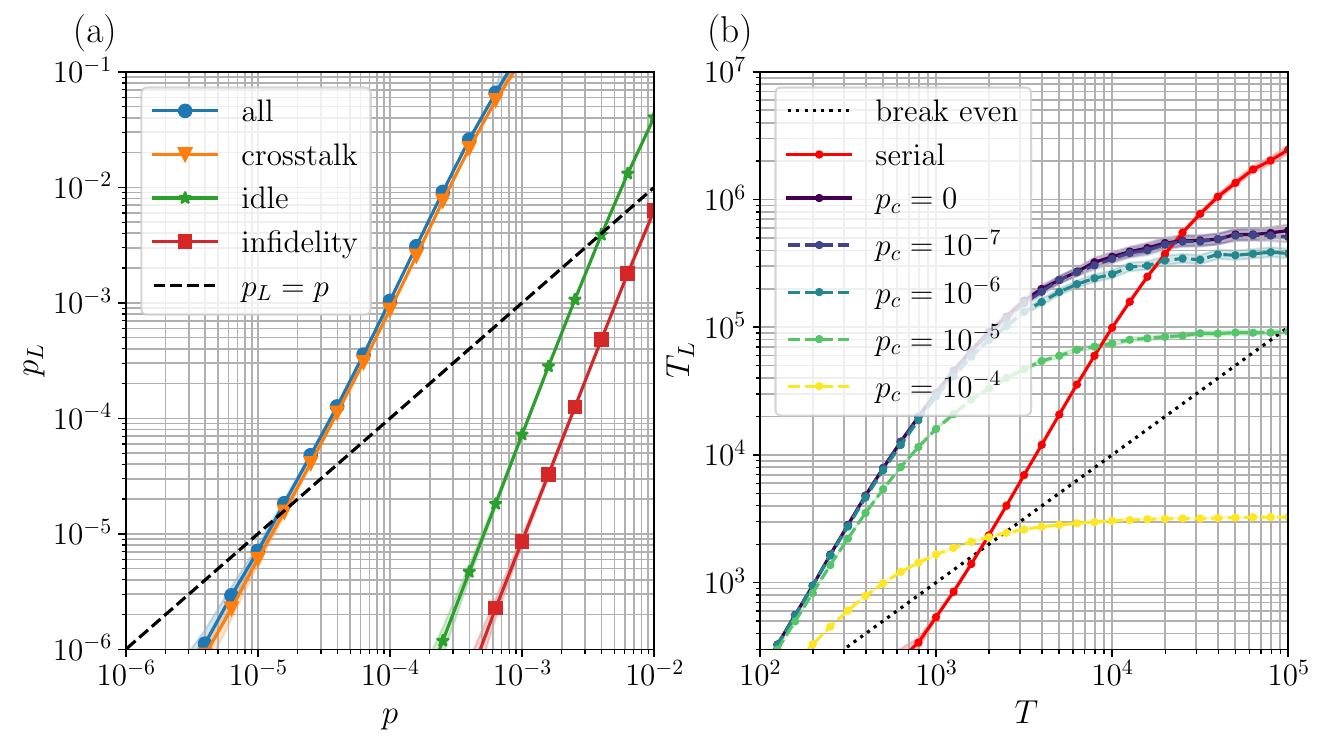}\\
    \caption{(a) The logical error rate for the $d=5$ code under different types of physical errors. Due to the two-qubit nature of the crosstalk error and the larger number of gate pairs to be applied in parallel, the logical error due to the crosstalk ($\propto p^2$) is much larger than those due to the two-qubit gate infidelity and the idling error ($\propto p^3$).
    (b) The logical coherence time $T_L$ versus the physical coherence time $T$ under a fixed two-qubit gate infidelity $p_g=10^{-3}$ and different levels of crosstalk errors $p_c$. The coherence times are in the unit of the two-qubit gate time. The dotted line indicate the break-even point $T_L=T$, and the red curve is when the entangling gates are applied in serial. \label{fig:surfacefull}}
\end{figure*}

To further visualize the significance of the crosstalk error in ion trap QEC with parallel gates, in Fig.~\ref{fig:surfacefull}(a) we compare the individual contributions of the three types of physical errors to the logical error. As we can see, the curves for the gate infidelity (red) and the idling error (green) are far below that for the crosstalk error (orange) with different scalings: While the curve for the crosstalk error scales as $p_L \propto p^2$, those for other physical errors scale as $p_L \propto p^3$ which is the expected behavior of the $d=5$ surface code. In this sense, we conclude that under similar values, the effect of the crosstalk error is much more significant than those of the gate infidelity and the idling error. Fortunately, later we will see that under typical noise levels, the magnitude of the crosstalk error is usually much smaller than the gate infidelity and the idling error, leaving us some room for balancing their contributions and optimizing the overall logical error rate.

Another widely concerned milestone for QEC is the break-even point where the logical coherence time reaches the physical coherence time. Here the logical coherence time can be defined in a similar way as Eq.~(\ref{eq:coherence}) with the small modification that the factor of $4/3$ for the depolarizing error is replaced by $2$ for the bit-flip error of the initial logical $|0_L\rangle$ state. Also we replace the time unit $t_0$ by the total duration of each syndrome measurement cycle. In Fig.~\ref{fig:surfacefull}(b), we plot the logical coherence time $T_L$ of the $d=5$ code versus the physical coherence time $T$ under a gate infidelity $p_g = 10^{-3}$ and various crosstalk error rates. For sufficiently small crosstalk error rates, e.g. $p_c< 10^{-6}$, we observe that the curves converge to the $p_c=0$ case, meaning that the logical error is dominated by the gate infidelity and the idling error. As $p_c$ increases, $T_L$ generally decreases but can still be above the break-even point $T_L=T$ (dotted line) for a wide range of parameters when $T$ is short. For longer $T$ and larger $p_c$, it will be more advantageous to apply entangling gates in serial as shown by the red curve. Note that in the regime of long physical coherence time $T$, the logical coherence time for serial gates is above that for parallel gates even when $p_c=0$. This is because, in this regime the logical error rates are dominated by the gate infidelity, which is the same for the parallel and serial cases. However, for serial execution, the time for each round of syndrome measurement is longer, leading to longer effective logical coherence time.

\begin{figure*}[!tbp]
    \centering
    \includegraphics[width=0.8\linewidth]{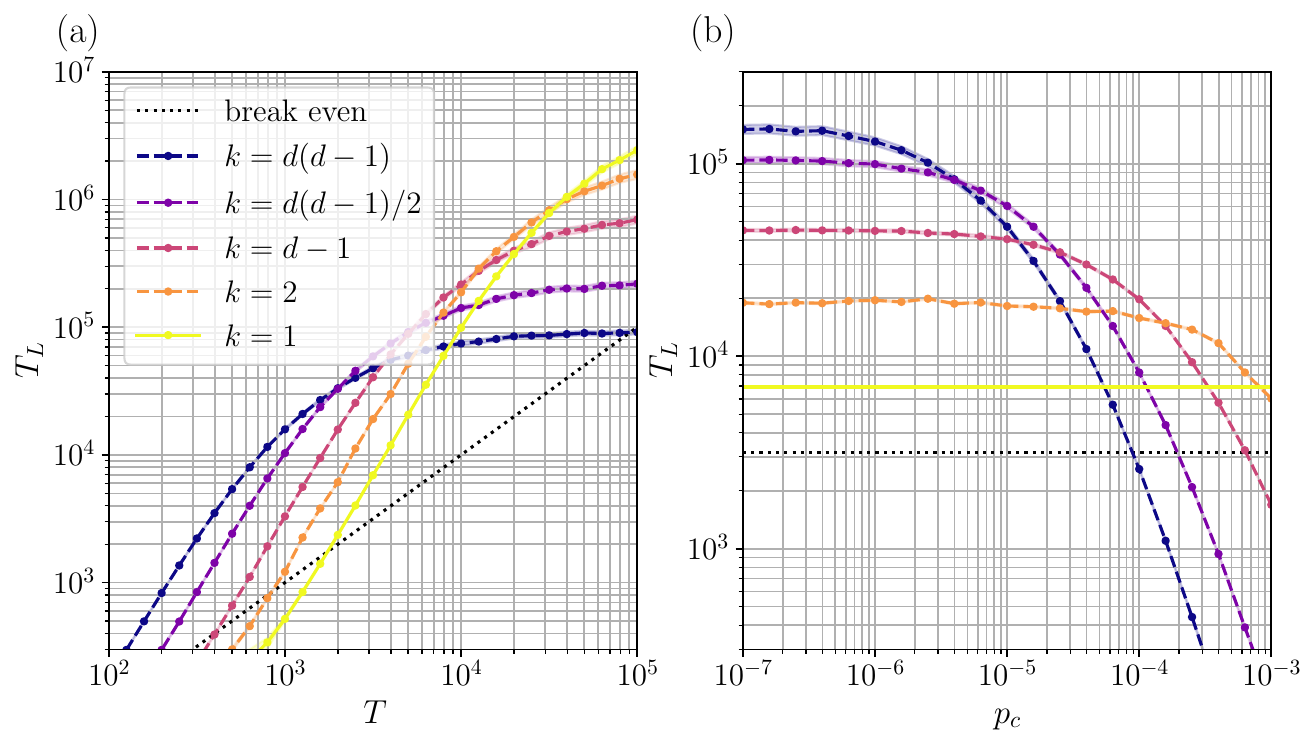}\\
    \caption{(a) The logical coherence time $T_L$ of the $d=5$ code versus the physical coherence time $T$ under a fixed two-qubit gate infidelity $p_g=10^{-3}$, a fixed crosstalk error rate $p_c=10^{-5}$, and different levels of parallelism. We use $k$ to represent the number of gate pairs to be applied in parallel. The coherence times are in the unit of the two-qubit gate time. At low $T$ it is more advantageous to parallelize the gates, while at high $T$ it is preferable to apply gates in serial. For intermediate parameters, an optimal parallelism level can be found due to the competition between the crosstalk error and the idling error. (b) The logical coherence time $T_L$ versus the crosstalk error rate $p_c$ under a fixed two-qubit gate infidelity $p_g=10^{-3}$, a fixed physical coherence time $T=10^{3.5}$, and different levels of parallelism. Under weaker $p_c$ it is more advantageous to execute gates in parallel while stronger $p_c$ prefers serial gates. Nevertheless, for a wide range of parameters, the break-even point $T_L=T$ (dotted line) can be surpassed. \label{fig:surfacepartial}}
\end{figure*}

\section{Optimizing parallelism level}
\label{sec:optimal_parallelism}
From the above discussion, we see that the advantage of parallelism is governed by the relative strength of different types of physical errors. In this section, we further optimize the parallelism level for different parameter regimes. Specifically, we fix a gate infidelity $p_g=10^{-3}$, and examine the competition between the crosstalk error and the idling error.

As described in Sec.~\ref{sec:code}, each round of syndrome measurement can be divided into $4$ layers of CNOT gates under maximal parallelism, where each layer contains $d(d-1)$ pairs of gates. Here we further assume that we apply $k$ pairs of gates in parallel, with $k=d(d-1)$ being the full parallelism case and $k=1$ being the serial case. In Fig.~\ref{fig:surfacepartial}(a) we plot the logical coherence time $T_L$ versus the physical coherence time $T$ under a fixed crosstalk error rate $p_c = 10^{-5}$ and different parallelism levels. As we can see, for a short physical coherence time (say, below $10^3$ CNOT gate time), it is more advantageous to fully parallelize the entangling gates and the shortened cycle duration can prevail the increased crosstalk error. On the other hand, for a very long physical coherence time (say, above $10^5$ CNOT gate time), for the $d=5$ code it is preferable to perform the entangling gates in serial while the idling error is still moderate. In the middle of these two extreme cases, we can design an optimal parallelism level to maximize the logical coherence time, exceeding the break-even point $T_L=T$ for a wide range of parameters.

Similarly, in Fig.~\ref{fig:surfacepartial}(b) we fix the physical coherence time to be $T = 10^{3.5}$ CNOT gate time, and scan the crosstalk error rate $p_c$. Here it is desirable to fully parallelize the entangling gates when the crosstalk is extremely low, say, below $5\times 10^{-6}$, while it is better to serialize gates for sufficiently high crosstalk error rate, say, above $10^{-3}$. Again, in the middle we can find an optimal parallelism level due to the competition between the crosstalk and the idling error, and the break-even point can be reached for a wide range of parameters.

\begin{figure}[!tbp]
    \centering
    \includegraphics[width=\linewidth]{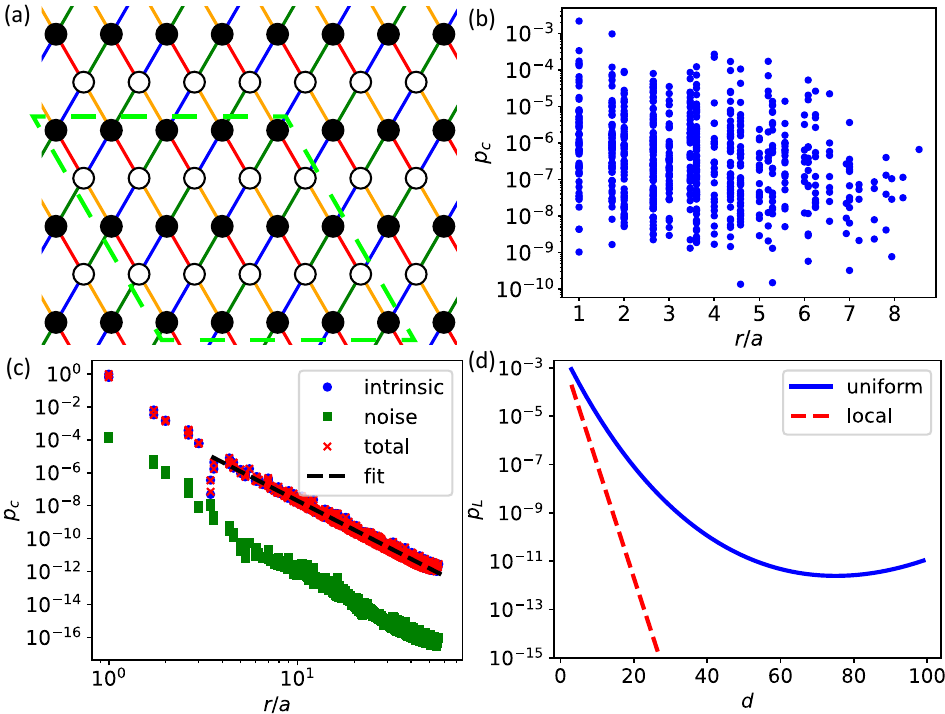}\\
    \caption{Scalability of ion trap quantum error correction under crosstalk errors. (a) We embed the rotated surface code onto a triangular lattice with a lattice constant of $a$. White (black) dots represent data (ancilla) qubits. Each color of edges represents a group of two-qubit gates that can be fully parallelized in (b). The dashed box represents a further division into sublattices in (c) with totally $8l^2$ parallel groups ($l=2$ illustrated here).
    (b) We design $d(d-1)=20$ pairs of parallel gates colored in red in (a) in a $d=5$ code using 500 segments and a total gate time $\tau=500\,\upmu$s under a practical ion spacing $a=5\,\upmu$m. We further assume independent amplitude fluctuation in each segment within $\pm1\%$ and a frequency shift within $\pm2\pi\times0.5\,$kHz to evaluate the crosstalk error between all pairs. The crosstalk shows no significant dependence on the distance $r$ between the ion pairs.
    (c) For a larger ion spacing $a=8\,\upmu$m, phonon modes involved in nearest-neighbor entangling gates are largely localized. We design a pulse sequence for one ion pair and apply the same sequence to all the parallel gates. The intrinsic crosstalk error due to this simplified pulse sequence and the additional crosstalk error due to the parameter fluctuation both scale polynomially with the distance $r$.
    (d) For the spatially uniform crosstalk in (b), we choose an optimized parallelism level $k=d-1$ such that both the crosstalk and the idling error scale as $O(d)$. Assuming a gate infidelity $p_g=3\times 10^{-3}$, a crosstalk $p_c=10^{-5}$ and a coherence time $T=5\times 10^4$, we obtain the logical error rate $p_L$ as the blue curve versus the code distance $d$. For the local crosstalk in (c), we choose a sublattice size $l=4$ to obtain a total crosstalk per gate $\tilde{p}_c=10^{-6}$. For a gate infidelity $p_g=10^{-3}$ and a coherence time $T=10^5$, we get an exponential decay of $p_L$ vs. $d$ as the red dashed line.
   \label{fig:scaling}}
\end{figure}

\section{Scaling of the logical error rate versus code distance}
\label{sec:scaling}
After examining the effect of the crosstalk error and its competition with the idling error in gate parallelism for a $d=5$ code, now we further discuss the scalability of ion trap QEC under the crosstalk error when increasing the code distance $d$. As shown in Appendix~\ref{sec:appendix}, we can use a unified expression
\begin{equation}
p_L\le 0.015[(p_g + 3t/8T + 1.3\tilde{p}_c)/0.013]^{(d+1)/2} \label{eq:scaling}
\end{equation}
to bound the total effect of the three types of physical errors on the logical error rate, where $t$ is the duration of each round of syndrome measurement, $T$ is the physical coherence time of the qubits, and $\tilde{p}_c$ is the average crosstalk error per targeted entangling gate. Suppose we parallelize $k$ pairs of entangling gates, we have $t\approx 4d(d-1)/k+5$ where we have neglected the faster state initialization and single-qubit gate operations, and the last term of $5$ is the measurement time. To estimate $\tilde{p}_c$ and its scaling with the code distance $d$ and the parallelism level $k$, we need to specify the physical arrangement of the ion crystal and the laser pulse sequence of the parallel gates. As shown in Fig.~\ref{fig:scaling}(a), we embed the rotated surface code onto a 2D triangular lattice of trapped ${}^{171}\mathrm{Yb}^+$ ions. We assume a practical transverse trap frequency $\omega_x=2\pi\times 3\,$MHz perpendicular to the plane, and we use counter-propagating $355\,$nm pulsed laser beams to address the qubits by Raman transitions \cite{Hou2024}. The edges represent the entangling gates between nearest-neighbor data qubits and ancilla qubits for the syndrome measurement. We use four colors for the edges to represent the four layers of entangling gates in Fig.~\ref{fig:surface}(b) that can be fully parallelized.

First, we consider a realistic ion spacing $a=5\,\upmu$m, and design all the required parallel gates for a $d=5$ code (with totally $2d^2-1=49$ physical qubits and four layers of parallel gates each containing $d(d-1)=20$ gate pairs) by the EASE algorithm \cite{Grzesiak2020} (see Appendix~\ref{sec:EASE} for details). As an example, we consider the gate pairs colored in red in Fig.~\ref{fig:scaling}(a) and design the pulse sequence using $500$ segments and a total gate time of $500\,\upmu$s. The complete pulse sequence can be found in Ref.~\cite{data_crosstalk} with a maximal Rabi frequency of $2\pi\times 385\,$kHz and an average Rabi frequency of $2\pi\times 66\,$kHz. By rotation symmetry, the same pulse sequence also works for the other three parallel gate groups.

Ideally, these parallel gates are designed to completely suppress the crosstalk error. However, due to the miscalibration of parameters in the experiment, such a cancellation is not perfect. In Fig.~\ref{fig:scaling}(b) we randomly sample an independent amplitude fluctuation for each ion and each segment which distributes uniformly in $\pm 1\%$, and a frequency shift of the laser which distributes uniformly in $\pm 2\pi\times 0.5\,$kHz. For each sampled parameter shift, we compute the actual two-qubit phase $\Theta_{ij}$ between all the undesired ion pairs and obtain the crosstalk error as $p_c=\Theta_{ij}^2$. We further average over 100 independently generated samples to get the crosstalk errors in Fig.~\ref{fig:scaling}(b) versus the distance $r$ between the ion pairs. As we can see, the crosstalk error shows no significant spatial dependence, and we obtain an average crosstalk $p_c\approx 1.1\times 10^{-5}$ for all the pairs. Also note that, under the same noise level, the gate infidelity for these individual gates already reaches $5\times 10^{-3}$. This suggests that under typical noise levels, the magnitude of the crosstalk error should be much smaller than that of the gate infidelity.

Based on the above simulation results, we model a uniform crosstalk error $p_c=10^{-5}$ between all the ion pairs. In this case, we have $\tilde{p}_c=2(k-1)p_c$ since there are $2k(k-1)$ possible locations of two-qubit crosstalk errors when parallelizing $k$ pairs of gates. Then in Eq.~(\ref{eq:scaling}) we have $O(d^2/k)$ idling error and $O(k)$ crosstalk error, hence an optimal parallelism level can be chosen as $k=O(d)$.
Note that Eq.~(\ref{eq:scaling}) suggests that in this parameter regime, an error threshold does not exist in the sense that for any finite $T$ and nonzero $p_c$ the logical error does not go to zero in the limit $d\to \infty$. Nevertheless, under reasonable parameters and a suitably choice of the code distance $d$, a sufficiently low logical error rate, e.g. below $10^{-10}$ can still be achieved which suffices for practical applications \cite{PhysRevA.86.032324,Acharya2024}. For example, in Fig.~\ref{fig:scaling}(d) we set $p_g=3\times 10^{-3}$ and $T=5\times 10^4$, and obtain the scaling of the logical error rate $p_L$ versus the code distance $d$ as the blue solid curve. We can choose $d=41$, requiring a 2D crystal of thousands of ions, to reach the logical error rate of $10^{-10}$.

The spatially uniform crosstalk error can be understood from a comparison between the gate speed and the speed of the sound wave on the 2D crystal. For a large ion crystal, the collective oscillation modes in the transverse direction can be decomposed into traveling waves in different directions \cite{PhysRevA.100.022332,Wu_2020}, with a group velocity proportional to $\epsilon \omega_x a$ where $\epsilon=e^2/(4\pi\epsilon_0 m \omega_x^2 a^3)$ is a dimensionless parameter characterizing the bandwidth of the transverse phonon modes, $a$ is the lattice constant and $m$ is the mass of the ions. When the gate speed is much slower than $1/(\epsilon \omega_x)$, we can expect that each target ion is effectively interacting with all the ions reached by the wavefront of the sound wave, resulting in a long-range crosstalk error. For the above example with $a=5\,\upmu$m, we have $\epsilon=0.0183$, and the sound wave from one site can propagate to a radius of hundreds of sites during the gate time $\tau=500\,\upmu$s, already covering the $d=5$ lattice considered above. Actually, as the ions are effectively interacting with more ions, the design of the pulse sequence also becomes more complicated and requires a longer time and more segments to disentangle all the phonon modes, which in turn leads to the propagation of the interaction to even farther away ions.

Given the above understanding, we may attempt to move to the ``fast gate'' regime to obtain a more preferable spatially localized crosstalk error. Note that here by ``fast gate'' we do not require the gate speed to be below or comparable to $1/\omega_x$, which is the topic of many previous literatures based on spin-dependent momentum kicks \cite{PhysRevLett.91.157901,PhysRevLett.93.100502,PhysRevLett.119.230501,PhysRevLett.120.220501,Wu_2020}. Instead, here we slightly increase the ion spacing to $a=8\,\upmu$m to reach $\epsilon=0.0045$, and use a moderate gate time $\tau=100\,\upmu$s to limit the propagation to just a few sites. Then we can design the pulse sequence for a high-fidelity entangling gate between a central ion pair in a $d=11$ lattice using the standard amplitude modulation method \cite{Zhu2006}, and apply the same sequence to all the targeted parallel gate pairs (we rescale the laser amplitude for different gate pairs slightly to compensate the boundary effect to get an ideal two-qubit phase $\Theta_{ij}=\pm \pi/4$). The complete pulse sequence together with the rescaling factor for different gate pairs in a larger $d=31$ code lattice can be found in Ref.~\cite{data_crosstalk}. Then again we compute the crosstalk error between the undesired ion pairs in Fig.~\ref{fig:scaling}(c) assuming $\pm 1\%$ amplitude fluctuation in each segment, and $\pm 2\pi\times 0.5\,$kHz frequency shift of the laser. However, note that in this case apart from such crosstalk errors due to the experimental noise, we further have an intrinsic crosstalk because of the use of a simplified pulse sequence which does not set $\Theta_{ij}$ to zero for undesired ion pairs even in the ideal case. We plot the intrinsic crosstalk error as $p_c^{(\mathrm{intrinsic})}=\Theta_{ij}^2$ without the sampled noise, and generate 100 samples of pulse noise to compute the standard deviation $\delta\Theta_{ij}$ and the crosstalk error due to noise as $p_c^{(\mathrm{noise})}=\delta\Theta_{ij}^2$. As we can see, both types of errors decay polynomially with the distance $r$ between the ion pairs, verifying the expected spatial locality in the ``fast gate'' regime. We fit a total crosstalk between one ion pair at a distance of $r$ as $p_c=0.015(r/a)^{-5.87}$ for $r\ge 4a$, indicated by the black dashed line.

This local crosstalk error model suggests that the average crosstalk error per targeted entangling gate $\tilde{p}_c$ can converge to a finite number even as $d\to \infty$. However, the price we pay is that we are not able to parallelize entangling gates close to each other because their intrinsic crosstalk is too large. As shown by the green dashed box in Fig.~\ref{fig:scaling}(a), we can divide the whole triangular lattice into sublattices with a lattice constant $2la$ ($l=1,2,3,\cdots$). This divides all the entangling gates into $8l^2$ subgroups to be parallelized, represented by the edges in the box. The sublattice structure ensures that the parallel gate pairs are separated by a distance of at least $(2l-1)a$, upper-bounding the crosstalk between them. Specifically, for $l=4$ we can compute the crosstalk error between all the parallel gate pairs using the above fitted model, and obtain $\tilde{p}_c=10^{-6}$. Also we get $t=8l^2+5$ for the $8l^2$ layers of parallel gates together with the measurement of the ancilla qubits, independent of the code distance $d$. Further assuming a gate infidelity $p_g=10^{-3}$ and a physical coherence time $T=10^5$, we obtain the red dashed line in Fig.~\ref{fig:scaling}(d). Here $d=17$ is sufficient to reach a logical error rate below $10^{-10}$, requiring a 2D crystal of hundreds of ions.

\section{Conclusion and outlook}
To sum up, in this work we show that the crosstalk error between parallel entangling gates in a large 2D ion crystal is an important consideration for quantum error correction. By its nature, the crosstalk error is a two-qubit error and requires a minimal code distance $d=5$ to correct it. Through an example of a $d=5$ rotated surface code, we show that the pseudo-threshold and the break-even point can be reached under realistic experimental parameters, and we discuss how the parallelism level can be optimized according to the competition between the crosstalk error and the idling error. We further study the spatial dependence of the crosstalk error in two different parameter regimes characterized by how the gate speed compares to the propagation speed of local oscillation on the 2D ion crystal. In the ``slow gate'' regime, the design of the parallel gates requires considering all the ions, leading to a long-range crosstalk error which only weakly depends on the distance between the parallel gates. In the ``fast gate'' regime, the design of the parallel gates can be simplified to include only a small fraction of ions that can be reached by the sound wave, and we obtain a crosstalk which decays polynomially with the distance between gate pairs. We discuss how the logical error rate scales with the code distance based on a unified formula to describe the joint effect of the gate infidelity, the idling error and the crosstalk error, and we show that in both parameter regimes a logical error rate below $10^{-10}$ can be achieved using a 2D crystal with hundreds to thousands of ions.

Note that, although trapped ions are well-known for their long coherence time, with the current record above one hour for a single ion \cite{wang2021single} and hundreds of milliseconds for hundreds of ions \cite{PhysRevA.106.062617}, these performances are only achieved with respect to a stable phase reference like a microwave source. For laser-based quantum gates, the physical coherence time may actually be limited by the laser dephasing time, which is typically tens to hundreds of milliseconds \cite{PhysRevLett.125.150505,PhysRevApplied.19.014014}. (For magnetic-field-sensitive qubits, additional limitation may come from the magnetic field noise \cite{PRXQuantum.2.020343}.) Such a relatively short coherence time, about $10^2$-$10^3$ of the two-qubit gate time, is advantageous for the demonstration of the break-even point in the near-term, but is not desirable for the ultimate goal of reaching sufficiently low logical error rates. To solve this problem, it may be preferable to use co-propagating Raman laser beams to realize single-qubit gates, and apply the light shift gate \cite{milburn2000,leibfried2003experimental,PhysRevLett.117.060504,PhysRevLett.127.130505}, the phase-insensitive Molmer-Sorensen gate \cite{Lee_2005}, or the phase-sensitive one with additional single-qubit gates to cancel the dependence on the laser phase \cite{Lee_2005,PhysRevLett.117.060505} for two-qubit entangling gates.

Finally, because in this work we use the rotated surface code as the example to discuss the effect of crosstalk errors, we only use nearest-neighbor entangling gates on the 2D crystal. However, due to the long-range Coulomb interaction between the ions, long-range entangling gates can also be performed with similar analysis for the crosstalk error, in particular in the ``slow gate'' regime. This enables the application of our results in other QEC codes like the quantum low-density parity-check (LDPC) code \cite{gottesman2014fault,PRXQuantum.2.040101} which can be partially parallelized, and may be used in transversal logical CNOT gates between two logical qubits encoded in different regions of a large 2D crystal.

\begin{acknowledgments}
This work was supported by the Innovation Program for Quantum Science and Technology (Grant No. 2021ZD0301601), the Shanghai Qi Zhi Institute, the Tsinghua University Dushi program, the Tsinghua University Initiative Scientific Research Program, and the Ministry of Education of China through its fund to the IIIS.
\end{acknowledgments}

\appendix
\section{Analytical scaling of logical error rates}
\label{sec:appendix}
\begin{figure*}[!tbp]
    \centering
    \includegraphics[width=0.8\linewidth]{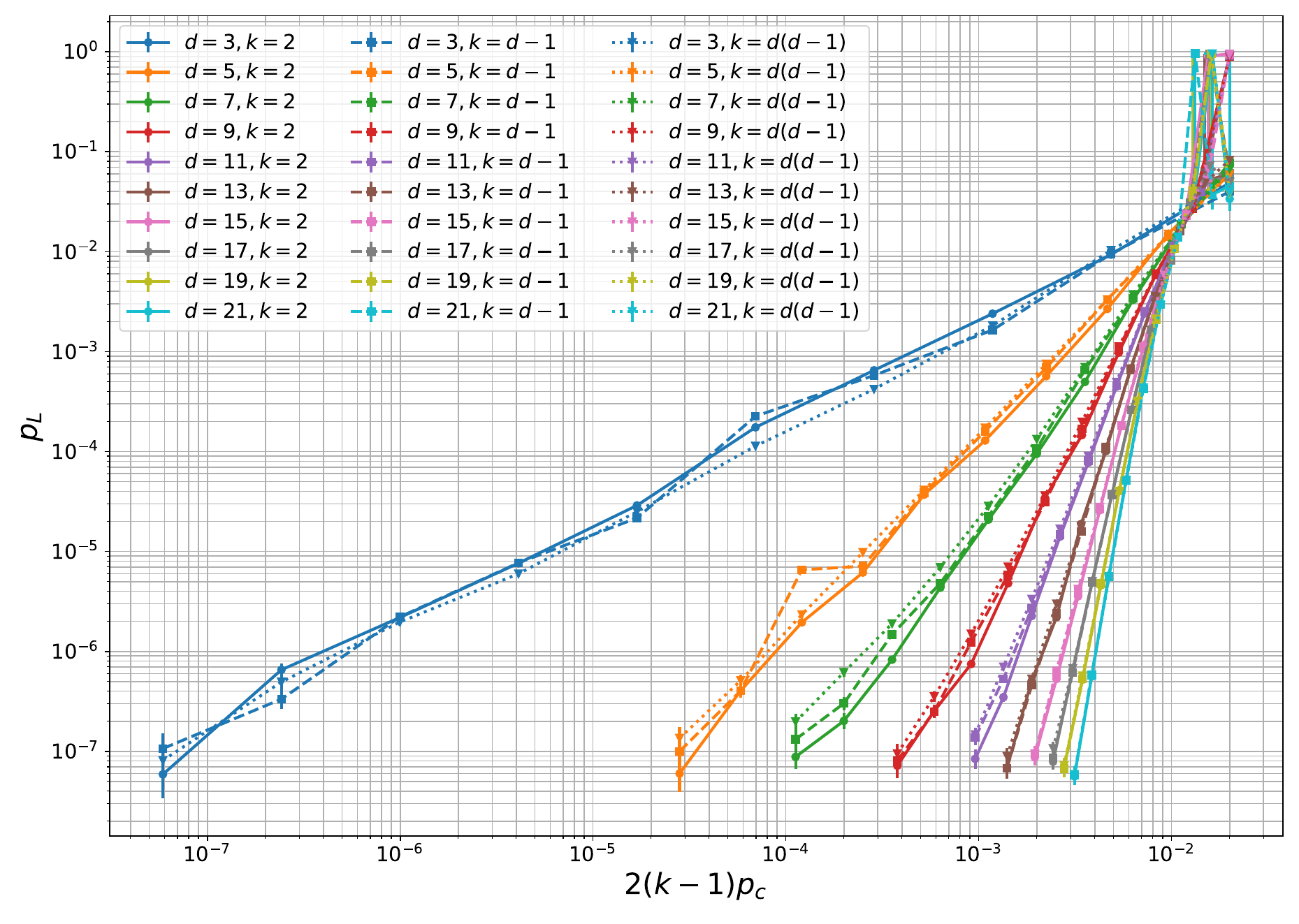}\\
    \caption{The logical error rate $p_L$ under different parallelism levels $k$ and different code distances $d$. The horizontal axis is scaled by $2(k-1)$ for different parallelism levels. To remove the internal structure in the errors for better scaling analysis, here different from the main text, we model the crosstalk error as a two-qubit depolarizing channel rather than the two-qubit Pauli $XX$ error. Also we randomize the order of the entangling gates when dividing them into subgroups for partial parallelism. \label{fig:crosstalk_total}}
\end{figure*}

\begin{figure}[!tbp]
    \centering
    \includegraphics[width=\linewidth]{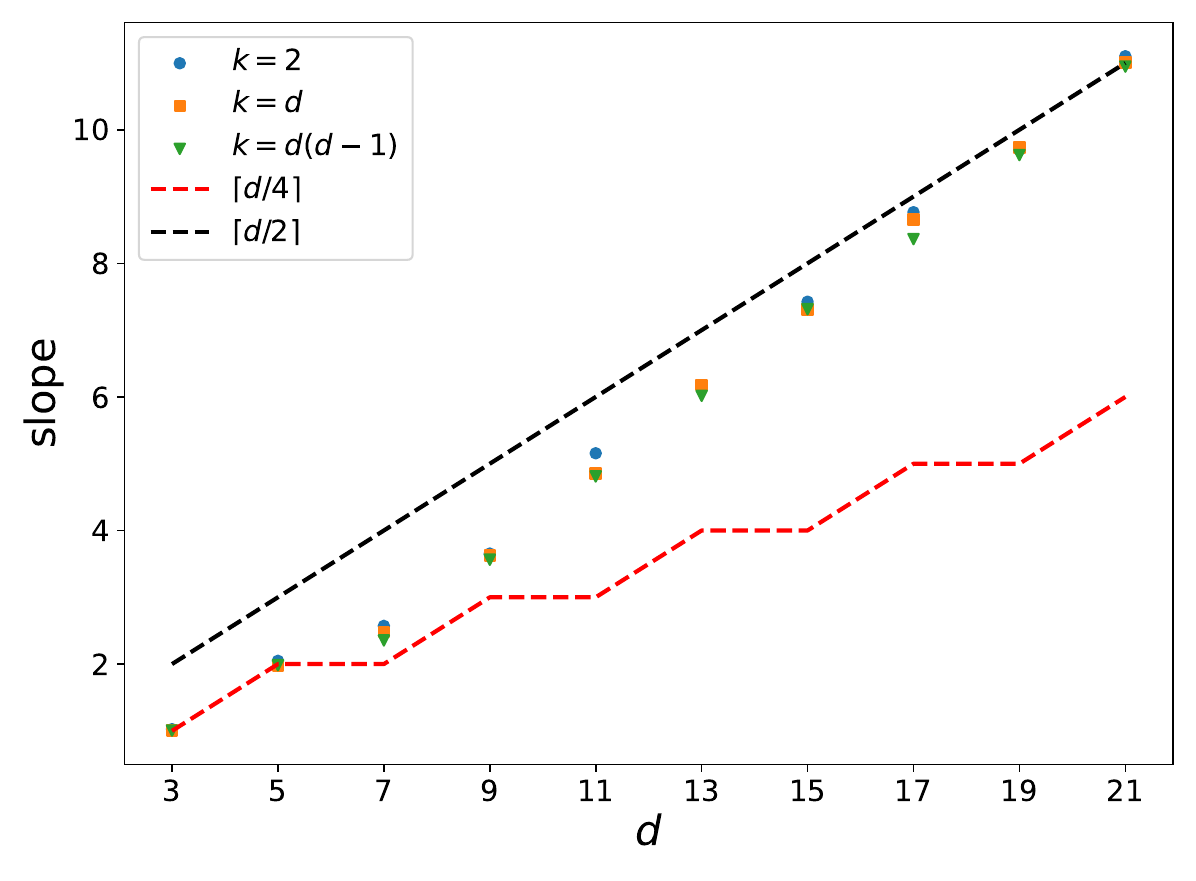}\\
    \caption{We plot the slope of the curves in Fig.~\ref{fig:crosstalk_total} near the error threshold under different parallelism levels $k$ and different code distances $d$. Theoretically, the asymptotic scaling approaches $\lceil d/4\rceil$ (red dashed curve) in the small $p_c$ limit. However, for practical crosstalk error rates near the error threshold and for large $d$, the scaling is closer to $\lceil d/2\rceil$ (black dashed curve). \label{fig:crosstalk_slope}}
\end{figure}

In Fig.~\ref{fig:crosstalk_total}, we simulate the logical error rate $p_L$ versus the crosstalk $p_c$ under different parallelism levels $k$ and different code distances $d$. Note that when applying $k$ entangling gates in parallel, the number of possible locations of two-qubit crosstalk is $2k(k-1)$. Assuming a constant crosstalk $p_c$ between all gate pairs, the crosstalk per entangling gate is $\tilde{p}_c=2(k-1)p_c$. Therefore we scale the horizontal axis for different parallelism levels by $2(k-1)$ accordingly. As we can see from the plot, the curves for different parallelism levels collapse well onto each other after this rescaling, which suggests that we can use a single number of the crosstalk per entangling gate $\tilde{p}_c$ to characterize the strength of the crosstalk error.

To further understand the scaling of the logical error rate versus the code distance $d$, in Fig.~\ref{fig:crosstalk_slope} we plot the slope of the curves in Fig.~\ref{fig:crosstalk_total} near the error threshold. For all the three types of parallelism levels, we see that the slope approaches $\lceil d/2\rceil$ as $d$ increases. This scaling may appear surprising, because by the two-qubit nature of the crosstalk error, we expect $p_L\sim p_c^{\lceil d/4\rceil}$ in the limit $p_c\to 0$. However, this asymptotic scaling may not hold for practical physical error rates close to the threshold \cite{PhysRevA.87.040301}. Here we provide a hand-waving argument for the observed scaling under full parallelism.

For simplicity, we consider surface codes with $d=4m$ ($m=1,2,\cdots$) such that we do not need to deal with the $\lceil\cdot\rceil$ function in the derivation. We assume an independent probability of $p_c$ to have two-qubit crosstalk errors between any qubit pairs that are involved in the $O(d^2)$ parallel gates. In the limit $p_c\to 0$, the most probable uncorrectable errors correspond to a shortest error chain with $d/2$ sites in the same row or the same column in Fig.~\ref{fig:surface}(a) (along the logical $X$ or $Z$ operators), or $d/4$ independent crosstalk errors. This corresponds to the above asymptotic scaling of $p_L\sim p_c^{d/4}$. If we count the total probability of such events, we have
\begin{equation}
p_{d/4} = 2d \frac{d!}{(d/2)!(d/2)!}\frac{(d/2)!}{(d/4)!2^{d/4}} p_c^{d/4},
\label{eq:shortest}
\end{equation}
where the first term represents the $2d$ possible rows or columns, the second term chooses the locations of the $d/2$ sites in the same row or column, and the third term divides these sites into $d/4$ ion pairs without considering their order.

On the other hand, the observed $p_L\sim p_c^{d/2}$ scaling suggests that we consider events with $d/2$ independent crosstalk errors.
These errors involve $d$ sites randomly chosen from the total $O(d^2)$ qubits.
%so the number of collisions in these $d$ sites is $O(1) \ll d$ and can be ignored.
% expectation value of m*(1-((k-1)/k)^m) for choosing m sites from totally k
We compute the probability that $d/2$ of these sites are in the same row or the same column, while the rest $d/2$ locate elsewhere, which is given by
\begin{equation}
p_{d/2} = 2d \frac{d!}{(d/2)!(d/2)!} (d^2 p_c)^{d/2} \label{eq:typical},
\end{equation}
where the first two terms are the same as those for $p_{d/4}$ and the factor of $d^2$ in front of $p_c$ comes from the fact that each given qubit may have crosstalk with $O(d^2)$ qubits, which is basically the $\tilde{p}_c$ defined above. Note that this expression already looks similar to that for memory errors of the surface code \cite{PhysRevA.86.032324} after substituting the physical error rate $p$ with $\tilde{p}_c$, hence explaining the observed scaling of $p_L\propto (\tilde{p}_c / p_{th})^{d/2}$.

Finally we examine the condition where Eq.~(\ref{eq:typical}) dominates over Eq.~(\ref{eq:shortest}). Note that both equations are derived under full parallelism of $O(d^2)$ gates. By requiring $p_{d/2}>p_{d/4}$, we get $p_c d^3 > e^{-\ln 2 - 1}$, i.e. $p_c d^2 > (1/d)e^{-\ln 2 - 1}$. Although the detailed value on the right-hand-side may deviate from the observed threshold due to the rough estimation of parameters, this condition suggests that the parameter regime where the $p_c^{d/2}$ scaling dominates becomes larger as $d$ increases, which is consistent with the observation in Fig.~\ref{fig:crosstalk_slope} that the slope approaches $d/2$ as $d$ increases. \hfill$\square$

Based on the above numerical results, now we can fit $p_L=A(\tilde{p}_c/p_{\mathrm{th}})^{(d+1)/2}$ where $A\approx 0.01$, $p_{\mathrm{th}}\approx 0.01$, and $\tilde{p}_c$ is the average crosstalk error per targeted entangling gate. Similarly, for the two-qubit entangling gate with a depolarizing error $p_g$, we perform numerical simulation (not shown) and fit $p_L=B(p_g/p_{\mathrm{th}}^\prime)^{(d+1)/2}$ where $B\approx 0.015$ and $p_{\mathrm{th}}^\prime\approx 0.013$. As for the idling error, for a physical coherence time $T$ and a duration $t\ll T$ for each cycle, we can regard it as a single-qubit depolarizing error of $(3/4)(1-e^{-t/T})\approx 3t/4T$. Therefore, we can upper bound its effect by absorbing it into a gate error of $3t/4T \times 2/4=3t/8T$ for each entangling gate where an additional factor of $1/4$ comes from the four entangling gates acting on each physical qubit in a QEC cycle, and a factor of $2$ from the fact that each entangling gate involves two qubits.

Finally, we combine all these three types of error together into $p_L\le \max\{A,B\}\{[p_g + 3t/8T + (p_{\mathrm{th}}^\prime/p_{\mathrm{th}})\tilde{p}_c]/p_{\mathrm{th}}^\prime\}^{(d+1)/2}$. This gives the expression we use in the main text for the asymptotic scaling of the logical error under large $d$.

\section{EASE algorithm for designing parallel entangling gates}
\label{sec:EASE}
The Efficient Arbitrary Simultaneously Entangling (EASE) protocol is a method for designing parallel entangling gates by amplitude modulation \cite{Grzesiak2020}. It divides the total gate time into $n_{\mathrm{seg}}$ equal segments, and finds an $n_{\mathrm{seg}}$-dimensional vector $\boldsymbol{\Omega}_j$ for each ion $j$ to represent the laser amplitude on each segment. The target is to set the residual spin-dependent displacement $\alpha_{jk}$ to zero in Eq.~(\ref{eq:alpha}) and the two-qubit phase $\Theta_{ij}$ in Eq.~(\ref{eq:theta}) to the desired values $\theta_{ij}$. Here we briefly summarize the procedure of this algorithm for completeness.

Consider parallel gates between $N$ ions in a crystal with $K$ collective phonon modes. For the piecewise-constant laser amplitudes defined above, we can integrate Eq.~(\ref{eq:alpha}) and Eq.~(\ref{eq:theta}) over each segment to express the conditions for the ideal gate design as
\begin{equation}
    A \boldsymbol{\Omega}_j = 0 \quad(\forall j), \label{eq:A_constraint}
\end{equation}
and
\begin{equation}
    \boldsymbol{\Omega}_i^T M_{ij} \boldsymbol{\Omega}_{j} = \theta_{ij} \quad(\forall i<j), \label{eq:M_constraint}
\end{equation}
where $A$ is a $K\times n_{\mathrm{seg}}$ matrix for the spin-dependent displacement, and $M_{ij}$ is an $n_{\mathrm{seg}}\times n_{\mathrm{seg}}$ matrix for the two-qubit entangling phase between the ion pair $i$ and $j$.

To enforce Eq.~(\ref{eq:A_constraint}), we restrict each $\boldsymbol{\Omega}_j$ to the null space of $A$ by organizing its orthonormal basis vectors into a matrix $\widetilde{A}$ and requiring
\begin{equation}
    \boldsymbol{\Omega}_j = \widetilde{A} \widetilde{\boldsymbol{\Omega}}_j,
    \label{eq:kernel_A_tranform}
\end{equation}
where $\widetilde{\boldsymbol{\Omega}}_j$ is a column vector with the same dimension as the null space of $A$, satisfying $\|\widetilde{\boldsymbol{\Omega}}_j\|_2 = \|\boldsymbol{\Omega}_j\|_2$.

By substituting Eq.~(\ref{eq:kernel_A_tranform}) into Eq.~(\ref{eq:M_constraint}), we transform the quadratic constraints into
\begin{equation}
    \widetilde{\boldsymbol{\Omega}}_i^T \widetilde{M}_{ij} \widetilde{\boldsymbol{\Omega}}_j = \theta_{ij} \quad(\forall i<j),
    \label{eq:M_constraint_transformed}
\end{equation}
where $\widetilde{M}_{ij} = \widetilde{A}^T M_{ij} \widetilde{A}$.

Now the EASE algorithm proceeds as follow. First we identify the connected components among the ions such that $\theta_{ij}=0$ for all the ion pairs located in different components. We reorder the ions to group the ions within the same component together, and to ensure that the first two ions in each component have nonzero targeted entanglement $\theta_{ij}$. (For the application in the main text, we have pairwise entanglement between the parallel gates, so either $\theta_{ij}=0$ or $\theta_{ij}=\pm\pi/4$, but here we describe the general algorithm to design arbitrary parallel gates.)
Then we sequentially determine $\widetilde{\boldsymbol{\Omega}}_{j}$ for each ion.

(1) For the first and the second ions in a connected component labeled by $k$ and $k+1$, we note that they are not entangled to any preceding ions, hence
\begin{equation}
    \widetilde{\boldsymbol{\Omega}}_{k^\prime}^T \widetilde{M}_{k^\prime,k} \widetilde{\boldsymbol{\Omega}}_{k}
    =
    \widetilde{\boldsymbol{\Omega}}_{k^\prime}^T \widetilde{M}_{k^\prime,k+1} \widetilde{\boldsymbol{\Omega}}_{k+1}
    = 0
    \quad (\forall k^\prime < k).
    \label{eq:pre-determined_first_two}
\end{equation}
We can construct a matrix for ion $k$
\begin{equation}
    \epsilon_{k}
    =
    \left[
    \begin{array}{c}
      \widetilde{\boldsymbol{\Omega}}_{1}^T \widetilde{M}_{1,k}\\
      \widetilde{\boldsymbol{\Omega}}_{2}^T \widetilde{M}_{2,k}\\
      \vdots \\
      \widetilde{\boldsymbol{\Omega}}_{k-1}^T \widetilde{M}_{k-1,k}
    \end{array}
    \right],
    \label{eq:epsilon_def}
\end{equation}
and similarly a matrix $\epsilon_{k+1}$ for ion $k+1$. Then the conditions in Eq.~(\ref{eq:pre-determined_first_two}) can be expressed as
\begin{equation}
    \epsilon_k \widetilde{\boldsymbol{\Omega}}_k
    =
    \epsilon_{k+1} \widetilde{\boldsymbol{\Omega}}_{k+1}
    =
    0,
\end{equation}
where again the solutions are restricted to the null space of these matrices.

The original EASE protocol determines $\widetilde{\boldsymbol{\Omega}}_{k}$ and $\widetilde{\boldsymbol{\Omega}}_{k+1}$ by finding the largest eigenvalue of $\widetilde{M}_{k,k+1}$ and projecting the corresponding eigenvector onto the null spaces of $\epsilon_{k}$ and $\epsilon_{k+1}$. Here we slightly modify this step to further lower the required laser intensity.

Similar as before, we use matrices $\widetilde{\epsilon}_k$ and $\widetilde{\epsilon}_{k+1}$ to denote the orthonormal basis vectors of the null spaces of $\epsilon_k$ and $\epsilon_{k+1}$, respectively. To restrict $\widetilde{\boldsymbol{\Omega}}_k$ and $\widetilde{\boldsymbol{\Omega}}_{k+1}$ to these null spaces, we require
\begin{equation}
    \widetilde{\boldsymbol{\Omega}}_{k}
    =
    \widetilde{\epsilon}_k \doublewidetilde{\boldsymbol{\Omega}}_k,
    \quad
    \widetilde{\boldsymbol{\Omega}}_{k+1}
    =
    \widetilde{\epsilon}_{k+1} \doublewidetilde{\boldsymbol{\Omega}}_{k+1}.
    \label{eq:kernel_epsilon_tranform}
\end{equation}
Then Eq.~(\ref{eq:M_constraint_transformed}) becomes
\begin{equation}
    \doublewidetilde{\boldsymbol{\Omega}}_k^T \doublewidetilde{M}_{k,k+1} \doublewidetilde{\boldsymbol{\Omega}}_{k+1} = \theta_{k,k+1},
\end{equation}
where $\doublewidetilde{M}_{k,k+1} = \widetilde{\epsilon}_k^T \widetilde{M}_{k,k+1} \widetilde{\epsilon}_{k+1}$. Now we get $\doublewidetilde{\boldsymbol{\Omega}}_k$ and $\doublewidetilde{\boldsymbol{\Omega}}_{k+1}$ as the left and right singular vectors of $\doublewidetilde{M}_{k,k+1}$ corresponding to the largest singular value, and rescale them to get the desired $\theta_{k,k+1}$. By plugging them back to Eq.~(\ref{eq:kernel_epsilon_tranform}) and Eq.~(\ref{eq:kernel_A_tranform}), we obtain the solutions to $\boldsymbol{\Omega}_{k}$.

(2) For a subsequent ion $k+l$ in a given connected component ($l\ge 2$), we similarly construct a matrix $\epsilon_{k+l}$ as Eq.~(\ref{eq:epsilon_def}) and require
\begin{equation}
\epsilon_{k+l} \widetilde{\boldsymbol{\Omega}}_{k+l} = [0,\,0,\,\cdots,\theta_{k,k+l},\,\cdots,\theta_{k+l-1,k+l}]^T.
\end{equation}
We can further minimize $\|\widetilde{\boldsymbol{\Omega}}_{k+l}\|_2$ subjected to this constraint, which is just the minimum-norm solution to a system of linear equations and can be computed using the standard linear algebra method. Note that here we mention this part of the algorithm for completeness, although in the main text we only consider pair-wise entangling gates so that the current situation will not occur.

%\bibliography{reference}

\end{document}